\documentclass[3p]{elsarticle}
\usepackage{epsfig}
\usepackage{amsfonts}
\usepackage{amsmath}
\usepackage{amssymb}
\usepackage{graphics}
\usepackage{mathrsfs}
\usepackage{color}
\usepackage{subfigure}
\usepackage{setspace}
\usepackage{graphicx}
\usepackage{mathrsfs}

\newtheorem{theorem}{Theorem}[section]

\newtheorem{example}{Example}[section]

\newtheorem{lemma}{Lemma}[section]

\newtheorem{definition}{Definition}[section]

\newproof{pol}{\bf{Proof of Lemma}}
\newproof{pot}{\bf{Proof}}

\journal{Communications in Nonlinear Science and Numerical Simulation}\begin{document}

\begin{frontmatter}
\title{\bf Periodic motions generated  from  non-autonomous grazing dynamics}

\author{M. U. Akhmet$^{*,a}$} \ead{marat@metu.edu.tr}
\author{ A.~K{\i}v{\i}lc{\i}m$^{a}$} \ead{kivilcim@metu.edu.tr}

\address{$^a$Department of Mathematics, Middle East Technical University, 06531 Ankara, Turkey}

\begin{abstract}
 This paper examines  impulsive non-autonomous systems with  grazing periodic solutions.  Surfaces of discontinuity   and impact  functions  of the  systems  are  not   depending on  the time  variable.   That   is,   we can  say  that  the impact  conditions  are  stationary,  and this   makes necessity  to   study the  problem  in  a new way. The   models    play   exceptionally  important  role in  mechanics   and electronics.    A concise review on the sufficient conditions for the new type of linearization  is presented. The existence and stability of   periodic   solutions  are considered  under the circumstance of the regular perturbation.  To visualize the theoretical results,  mechanical examples  are  presented.    
\end{abstract}


\end{frontmatter}

\section{Introduction} 
Impacting systems serve various types of dynamical properties such as chaos \cite{prii1,prii2} and chattering \cite{nordprii1,kry11}. One of the important phenomenon that occurs in the impacting systems is   grazing.  There exist many papers which consider the dynamics around the grazing point, the existence of the periodic solutions with grazing point and the complexity around the grazing point \cite{Bernardo-Hogan2010}-\cite{winston}. 
 
Dynamics with grazing points is complex and difficult for the analysis. There is a seminal method issuing from the papers \cite{feigin70,Nord1} is widely used for the investigation of such kind of problems. This method is based on a special map which is constructed as composition of several continuous and discrete motions. For our investigation, we utilized an approach which was summarized in \cite{ref1}.  For the analysis of grazing, the B-equivalence method recently started to be applied \cite{akh-kivi,akh-kivi2}. It is urgent to mention that diversity    of methods   determine  the   strength   of   mathematics. For this reason, we hope   that   the application of B-equivalent differential equations with impulses can enrich  results  for the grazing phenomenon.

In the study \cite{budd1}, a new form of bifurcation called the grazing bifurcation, which gives a rise to  complex dynamics including chaotic behavior interspersed with period adding windows of periodic behaviour, is identified and exemplified. The normal form for the grazing bifurcation is constructed to classify the dynamics around it. It is demonstrated that complex dynamic behavior can be found at a grazing bifurcation. In the studies  \cite{I1,I2}, the bifurcation around the grazing solution of the system is examined.  The parameter variation is observed only in the vector-field of an impacting system and some sufficient conditions for the stability of a grazing periodic solution of that system is presented. In the paper \cite{I1}, it is asserted that the grazing impact which is known to be a discontinuous bifurcation can be regularized with appropriate impact rule which differs in many aspects from the existing ones.  In \cite{I2}, considering the non-zero impact duration, the bifurcations which are related with grazing contacts are analyzed.  The theoretical results are exemplified by taking into account a mechanical system which consists of a disc with an offset center of gravity bouncing on an oscillating surface. These all papers are united in the following sense, vector-field is the function of both time and space variables and the surfaces of discontinuity and the jump operators of them are defined only through the space variables, then it is easy to call it a non-autonomous system with stationary impulses. However, there exist some systems where both the vector field and the surface of discontinuity consist of space variables, then they are called autonomous impulsive system \cite{brogliato}. Finally, there are papers about mechanical systems where vector field as well as surface of discontinuity defined by both time and space variables. Then, it is easy to see that such systems are non-autonomous impulsive systems \cite{ref1}.

For the analysis of such systems, it is significant to define the grazing properly. There exist two different approaches  in the literature for the definition of the grazing. One of them occurs when the impacting particle  touches with zero velocity  to the surface of discontinuity \cite{Nord1}-\cite{Nord2006}.  Another one take place when the particle meets the surface of discontinuity tangentially \cite{Bernardo-Hogan2010,budd2001,Luo2005a}. In the present paper, to develop theory for the  grazing systems with stationary impulses, we take into account the comprehension of the authors who assert that the solutions have a contact with  the surface of discontinuity tangentially at the grazing point.

The newly developed linearization  for  non-autonomous systems with stationary impacts  was considered in our paper \cite{akh-kivi2}.  We  investigate there   equations, which  are  non-perturbed,  while   the    present  research  is devoted to  extension of the   results to  more complex problems, when  grazing effects changes connected to  a parameter variations. 

 Due to the fact that the differential equation of the system is non-autonomous  but the impulse equation is autonomous, this system can be named ``half-autonomous system." In order to analyze these type of equations, we should request conditions which are in many senses different than those presented in  \cite{impactdef}. In other words, our present research is slightly different than that for autonomous impulsive systems.  

In the paper \cite{I1}, the grazing bifurcation is considered in impacting mechanical systems where the coefficient of restitution was proposed as a third order polynomial. By applying a special linearization technique the multipliers of the system has been obtained. However, to construct the linearization the relation between the surface of discontinuity and the vector field is not taken into account. In this situation, the grazing arises whenever the vector field is tangent to surface of discontinuity. For this reason, if one conduct a study about the stability of grazing solution, it is urgent to investigate the linearization around that solution by considering the surface of discontinuity.  In our paper \cite{akh-kivi}, we have considered the restitution coefficient as quadratic to suppress the grazing in the system and  in our analysis the surface  of discontinuity is considered  in the linearization. In the paper  \cite{I1}, the grazing bifurcation which leads the change in the stability of the periodic solution is taken into account. As different than these results, we have analyzed the existence of periodic solution under parameter variation where the stability is preserved. In this paper, we have consider the regular behavior around the grazing periodic solution of the system.

The remaining part of the paper is divided into four parts. The next section covers information of the half autonomous systems, grazing point and grazing solution and some sufficient conditions are provided. The third section is about the linearization of that system around the grazing periodic solution. The four section is related with the stability of  the periodic solution. In the fifth, the small parameter analysis have been conducted in the neighborhood of grazing periodic solution. Examples  of periodic  solutions   for  two  impact   mechanisms generated from models  with   grazing are  provided.  The last one is the discussion section which displays the sum of our work and possible future works related with our subject of discussion.

\section{The grazing solutions}

 Let  $\mathbb{R},$ $\mathbb{N}$ and $\mathbb{Z}$ be the sets of all real numbers, natural numbers and integers, respectively. Consider the open connected  and bounded   set $G\in\mathbb{R}^n.$ 
  Let  $\Phi: G \to  \mathbb R$ be a  function, differentiable up to second order with respect to $x,$
$S =  \Phi^{-1}(0)$  is  a closed subset of $G.$    Define  a  continuously  differentiable  function $J:G \rightarrow G $ such that $J(S)\subset G.$  The function $I(x)$ will be used in the following part of the paper which is defined as  $I(x):=J(x)-x,$ for $x\in S.$

The following definitions will be utilized in the remaining part of the paper. Let $x(\theta-)$ be the left limit of   a function $x(t)$ at the moment $\theta,$   and  $x(\theta+)$ be the right limit of the solution. Define  $\displaystyle{\Delta x(\theta):= x(\theta+)-x(\theta-)}$  as the jump operator for  $x(t)$ such that $x(\theta)\in S$ and $t=\theta$ is a moment of discontinuity. Discontinuity moments are the moments when the solution meets the surface of discontinuity.  

In this paper, we take into account the following  system

\begin{equation}\label{half-auto}
\begin{aligned}
&x^{\prime}=f(t,x),\\
&\Delta x|_{x\in S}=I(x),
\end{aligned}
\end{equation}
where $(t,x)\in \mathbb R\times G,$ the functions $f(t,x)$ is continuously differentiable with respect to $x$ up to second order and continuous with respect to time.   We will consider the surface of discontinuity as    $\Gamma =\{(t,x)| \Phi(x)=0\} \subseteq \mathbb{R}\times S.$  We say   that   the system is  with    stationary  impulse conditions,  since  the function  $I(x)$  and the surface $S$  do not  depend   on time.

For the convenience in notation,   let   us   separate   the  differential   equation of the  impulse system as 

\begin{equation}\label{diff}
\begin{aligned}
&y^{\prime}=f(t,y).
\end{aligned}
\end{equation}

Assume that the solution  $x_0(t)=x(t,t_0,x_0),$ $t_0\in \mathbb R,$ $x_0\in G$ of \eqref{half-auto} intersects the surface of discontinuity $\Gamma,$ at the moments $t=\theta_i, \ i\in\mathbb{Z}.$  

Set the gradient vector of $\Phi$ with respect to $x$ as $\nabla\Phi(x).$ The normal vector of $\Gamma$ at  a  meeting moment, $t=\theta_i,$ of the solution $x_0(t)$ can be determined as $\overrightarrow n=(0,\nabla\Phi(x_0(\theta_i)))\in\mathbb{R}^{n+1},$
 where  $\langle ,\rangle$   means the dot product.  For the tangency, the vectors $\overrightarrow n$ and $(1,f(\theta_i,x_0(\theta_i))) $ should be perpendicular. That is, $\langle \nabla\Phi(x_0(\theta_i)),f(\theta_i,x_0(\theta_i))\rangle=0.$ 
 
In what follows, let $\|\cdot\|$ be the Euclidean norm, that is for a vector $x=(x_1,x_2,\ldots,x_n)$ in $\mathbb{R}^n,$ the norm is equal to $\sqrt{x_1^2+x_2^2+\ldots +x_n^2}.$  

Consider the function $H(t,x):=\langle \nabla\Phi(x),f(t,x)\rangle,$ with $(t,x)\in\mathbb{R}\times S.$

 A point $(\theta_i,x_0(\theta_i))$    is a \textit{grazing one}   and $\theta_i$ \textit{ a grazing moment} for a solution $x_0(t)$ of (\ref{half-auto}) if  $H(\theta_i,x_0(\theta_i))=0,$   and  $I(x_0(\theta_i)) =0.$

 A solution $x_0(t)$ of (\ref{half-auto}) is  grazing  if it has a grazing point $(\theta_i,x_0(\theta_i)).$ 

A point $(\theta_i,x_0(\theta_i))$    is a \textit{transversal point}  and $\theta_i$ a transversal moment  for a solution $x_0(t)$ if  $H(\theta_i,x_0(\theta_i))\neq 0. $

In what follows, we will assume the validity of the following condition.

\begin{itemize}
\item[(H1)] For each grazing point $(\theta_i,x_0(\theta_i)),$ there is a number $\delta>0$ such that $H(t,x)\neq 0$ and $J(x)\notin S$ if $0<|t-\theta_i|<\delta$ and $0<\|x-x_0(\theta_i)\|<\delta.$
\end{itemize}

It is also   clear that  function $H(t,x)\neq 0$  near   a transversal point.

Consider a  solution $x(t)=x(t,\theta_i,x_0+\Delta x)$ of \eqref{half-auto}  with a small $\|\Delta x\|.$ Because of the geometrical reasons caused by the tangency at the grazing point,  this solution may not intersect the surface of discontinuity near $(\theta_i,x_0(\theta_i)).$  For this reason there exist two different behavior   of   it  with respect to the surface of discontinuity, they are:

\begin{itemize}
\item[(N1)] The solution $x(t)$ intersects the surface of discontinuity $\Gamma$    at a moment  near to $\theta_i.$ 
\item[(N2)] There is no  intersection moments   of  $x(t)$    close   to  $\theta_i.$   
\end{itemize}

We say that  $\theta=\{\theta_i\}$ is a \textit{$B-$sequence } if one of the following alternatives holds: $(i)$ $\theta=\emptyset,$ $(ii)$ $\theta$ is a nonempty and finite set, $(iii)$ $\theta$ is an infinite set such that $|\theta_i|\rightarrow\infty$ as $i\rightarrow\infty.$ In what follows, we will consider  $B-$sequences.

In order to define a solution of (\ref{half-auto}), the following functions and sets are needed.

A function $\phi(t):\mathbb{R}\rightarrow\mathbb{R}^{n}, \ n\in\mathbb{N},$ is piecewise continuously differentiable \cite{ref1}.

\subsection{B-equivalence  to  a system  with  fixed moments of impulses}\label{B-equivalence}
 In this subsection, we will construct a system with fixed moment of impulses  which preserves the dynamical properties of the system \eqref{half-auto} and it  is called a B-equivalent system to that \eqref{half-auto} \cite{ref1}.

Consider a solution $x_{0}(t): \mathscr{I}\rightarrow \mathbb{R}^n,$ $\mathscr{I}\subseteq \mathbb{R},$ of (\ref{half-auto}). Assume that all discontinuity points $\theta_i$   of   $x_0(t),$  $i\in\mathscr{A},$ are interior points of $\mathscr{I},$  where  $\mathscr{A}$  is an  interval in $\mathbb Z.$    There exists a positive number $r,$ such that $r$-neighborhoods  $G_i(r)$ of $(\theta_i ,x_0 (\theta_i))$ do not intersect each other. 
Fix $i\in\mathscr{A}$ and let $\xi(t)=x(t,\theta_i,x),$ $(\theta_i,x)\in G_i (r),$ be a solution  of  (\ref{diff}),  which    satisfies $(N1),$    and $\tau_i =\tau_i (x)$ the meeting time of $\xi(t)$ with $S$ and $\psi(t)=x(t,\tau_i,\xi(\tau_i)+J(\xi(\tau_i)))$ another solution  of  (\ref{diff}).  Denote $W_i(x)=\psi(\theta_i )-x$ and one can define the map $W_i (x)$ as
\begin{equation}\label{W-map}
W_i (x)= \int^{\tau _i}_{\theta _i}{f(s,\xi(s))ds}+J(x+\int^{\tau _i}_{\theta _i}{f(s,\xi(s))ds})+\int^{\theta _i}_{\tau _i}{f(s,\psi(s))ds}
\end{equation}
It  is a map of an intersection of the plane $t=\theta_i$ with $G_i (r)$ into the plane $t=\theta_i.$  If  $\xi(t)$  does not    intersect $S$  near 
$t= \theta_i,$ then $W_i (x) = 0.$  Let us present the following system of differential equations with impulses at fixed moments,

\begin{equation}\label{eq:graz2}
\begin{aligned}
&y'=f(t,y),\\
&\displaystyle{\Delta y|_{t=\theta_i }= W_i (y(\theta_i))}.
\end{aligned}
\end{equation}

The function $f$ is the same as the function in system (\ref{eq:graz2}) and the map $W_i,$ $i\in\mathscr{A},$ is defined by equation (\ref{W-map})  if $x(t)$  satisfies condition $(N1).$  Otherwise, 
if   a  solution  $x(t)$     satisfies  $(N2),$  then    we   assume   that it  admits  the   discontinuity   moment  $\theta_i$   with   zero  jump   such  that  $W_i(x(\theta_i)) =0.$

Let us introduce the sets $F_r =\{(t,x)|t\in \mathscr{I}, \|x-x_0 (t)\|<r\},$ and ${G}_i^+ (r),$ $i\in\mathscr{A},$ an $r-$ neighborhood of the point $(\theta_i ,x_0 (\theta_i+)).$ Write $G^r = F_1 \cup (\cup_{i\in\mathscr{A}} G_i (r))\cup (\cup_{i\in\mathscr{A}} {G}_i^+ (r)).$ Take $r$ sufficiently small so that $G^r \subset\mathbb{R}\times G.$ Denote by $G(h)$ an $h$-neighborhood of $x_0 (0).$

 Systems (\ref{half-auto}) and  (\ref{eq:graz2}) are said to be $B-$equivalent in  \index{$B-$equivalence}
$G^{r}$ if there exists  $h >0,$  such that:
\begin{enumerate}
\item for every solution $y(t)$ of (\ref{eq:graz2}) such that $y(0) \in G(h),$ the
integral curve of $y(t)$ belongs to $G^{r}$      there exists
a solution
   $x(t) = x(t,0,y(0))$ of    (\ref{half-auto}) which satisfies
\begin{eqnarray}
&&x(t) = y(t), \ \ t \in [a,b]\backslash \cup_{i=-k}^{m} (\widehat
{\tau_i,\theta_i]}, \label{e12}
\end{eqnarray}
where $\tau_i$ are moments of discontinuity of $x(t).$
Particularly:
\begin{equation}
\begin{array}{l}
x(\theta_i)  = \left \{\begin{array} {ll} y(\theta_i), \quad \mbox {\, if $ \theta_i \leq \tau_i$},\\
y(\theta_i+), \mbox {otherwise,} \end{array}\right.    \\ 
y(\tau_i)  = \left \{\begin{array} {ll} x(\tau_i), \quad \mbox {\, if $ \theta_i \geq\tau_i$},\\
x(\tau_i+), \mbox {otherwise.} \end{array}\right. \label{e13}
\end{array}
\end{equation}
\item Conversely, if  (\ref{half-auto}) has a solution
$x(t)=x(t,0,x(0)), x(0) \in G(h),$ then there exists a solution
$y(t)=y(t,0,x(0))$ of (\ref{eq:graz2}) which has an integral curve  in
$G^{r},$ and (\ref{e12})  holds. \label{deqv}
\end{enumerate}
\label{defneq}

The  solution $x_0(t)$ satisfies (\ref {half-auto}) and (\ref {eq:graz2})
simultaneously.

\section{Linearization around a grazing   solution}
In order to consider stability properties of any solution, we should consider the linearization system around that solution first.  For this reason, let us start with the linearization system around the  grazing solution $x_0(t)=x(t,0,x_0),\ x_0\in G,$ of (\ref{half-auto}) which was introduced in the last section. We will demonstrate that one can write the variational system for the solution  as follows:

\begin{equation}\label{lin1}
\begin{aligned}
&u^\prime=A(t)u, \\
&\Delta u|_{t=\theta_i} =B_i u(\theta_i),\\
\end{aligned} 
\end{equation} 
where the matrix $A(t)\in \mathbb{R}^{n\times n}$ of the form $A(t)=\frac{\partial f(t,x_0(t))}{\partial x}.$ We call the second  equation in (\ref{lin1}) as the \textit{linearization at a moment of discontinuity  or at   a point  of discontinuity}. It is different for transversal and grazing points. However, the first differential equation in (\ref{lin1}) is common for all type of solutions. The matrices  $B_i$ will be described in the remaining part of the paper for each type of  the   points. 

\subsection{Linearization at a transversal moment}\label{trans}

   Linearization at the transversal point   has been  analyzed  completely in  Chapter 6, \cite{ref1}. Let  us   demonstrate the   results  shortly. The $B-$ equivalent system (\ref{eq:graz2})   is   involved in the analysis,  since  the   solution $x_0(t)$  satisfies also  the   equation (\ref{eq:graz2})  at all moments of time, and near solutions do  the  same for  all moments except   small neighborhoods of the  discontinuity  moment $\theta_i.$   Consequently, it  is easy  to  see that     the system of variations    around   $x_0(t)$ for   \eqref{half-auto}  and (\ref{eq:graz2})  are   identical.  Assume that $x(\theta_i)$ is at a transversal point.  We consider   the   reduced  B-equivalent system      and use the  functions $\tau_i(x)$ and  $W_i(x),$ defined by equation (\ref{W-map}), are presented in Subsection \ref{B-equivalence} for   linearization.
Differentiating $\Phi(x(\tau_i(x)))=0,$ we have

\begin{equation}\label{dert}
\frac{\partial\tau_i(x_{0}(\theta_i))}{\partial x_{0j}}=-\frac{\langle\Phi_{x}(x_0(\theta_i)),\frac{\partial x_{0}(\theta_i)}{\partial x_{0j}}\rangle}{\langle\Phi_{x}(x_0(\theta_i)),f(\theta_i,x_0(\theta_i))\rangle}, j=1,\ldots,n.
\end{equation}

The  Jacobian $W_{ix}(x_0(\theta_i))=[\frac{\partial W_{i}(x_0(\theta_i))}{\partial x_{01}},\frac{\partial W_{i}(x_0(\theta_i))}{\partial x_{02}},\ldots,\frac{\partial W_{i}(x_0(\theta_i))}{\partial x_{0n}}]$  is   evaluated by 
\begin{equation}\label{derw}
\frac{\partial W_{i}(x_0(\theta_i))}{\partial x_{0j}}=(f(\theta_i,x_0(\theta_i))-f(\theta_i,x_0(\theta_i)+I(x_0(\theta_i))))\frac{\partial\tau_i}{\partial x_{0j}}+\frac{\partial I}{\partial x_0}(e_j+f(\theta_i,x_0(\theta_i))\frac{\partial\tau_i}{\partial x_{0j}}), 
\end{equation}
where $e_j=(\underbrace{0,\ldots,1}_j,\ldots,0),$ $ j=1,2,\ldots,n.$   Next,  by  considering the   second equation in  (\ref{eq:graz2}) and using mean value theorem,  one can obtain that 
$\Delta(x(\theta_i)-x_0(\theta_i))=W_i(x(\theta_i)-x_0(\theta_i))=W_{ix}(x_0(\theta_i))(x(\theta_i)-x_0(\theta_i))+O(\|x(\theta_i)-x_0(\theta_i)\|).$

From the   last   expression,  it is seen that   the linearization at the transversal moment  is   determined with the  matrix  $B_i=W_{ix}(x_0(\theta_i)).$

\subsection{Linearization at a grazing moment} \label{graz1}

Fix a grazing  moment  $\theta_i.$  Considering the definition  of grazing point with the formula  \eqref{dert}, it is apparent that at least  one coordinate of the gradient, $\nabla\tau(x),$  is infinity at the grazing point. This causes singularity in the system.  Through the formula \eqref{dert}, one can see that the singularity is just caused by the position of the vector field  with respect to the surface of discontinuity  and the impact does not participate in the appearance of the singularity.  To get rid of  the singularity, we will consider the following conditions.

\begin{itemize}
\item[$(A1)$] The   map  $W_i(x)$  in   \eqref{W-map} is   differentiable  if  $x = x_0(\theta_i).$  
\item[$(A2)$]    $\tau_i(x)<\theta_{i+1} -\theta_i-\epsilon$ for some positive $\epsilon$  on a set  of points near    $ x_0(\theta_i),$   which   satisfy   condition $(N1).$
\end{itemize}

The appearance of singularity in  \eqref{dert} does not mean that the Jacobian $W_{ix}(x)$ is infinity. Because, in order to find the Jacobian, not only the surface of discontinuity and the vector field are required, but the jump function is also needed. The regularity of the Jacobian  can be arranged  by means of the proper choice of the vector field, surface of discontinuity and jump function. In other words,  if they are specially chosen, the map can be differentiable, and   this validates condition $(A1).$   Thus,  in the present  paper  we analyze   the  case,   when   the impact  functions neutralize  the  singularity.     Presumably,   if there   is none of this  type  of  suppressing,  complex   dynamics near the  grazing  motions may  appear \cite{budd2001, Luo2006, Nord1,Nord97}.
In the examples stated in the remaining part of the paper, one can see the verification of $(A1),$ in details.

There are many ways are suggested to investigate the existence and stability of periodic solution of systems with graziness in literature \cite{Hosa-Champneys,Nord2001}. They investigate them by constructing special maps around the grazing point.  In this paper, we suggest to investigate the existence and stability  by using the method of Floquet multipliers  for   dynamics with continuous time.  It is a well known method in literature \cite{Hartmanode}, but this is not widely applied to the analysis of the stability of grazing solutions because of the tangency of the grazing solution with the surface of discontinuity. It constitutes  the main novelty of our paper.

By means of these discussions, one can conclude that  the matrix $B_i$ in \eqref{lin1} is the following

\begin{eqnarray}\label{grazBi}
B_i=\begin{cases}   W_{ix}\quad &\mbox{if } \quad \mbox{ $(N1)$ is valid,} \\ 
 O_{n} \quad &\mbox{if} \quad \mbox{ $(N2)$ is valid,} \end{cases}
\end{eqnarray}
where $O_{n}$ denotes the $n\times n$ zero matrix.

Denote by $\bar x(t), j=1,2,\ldots,n,$   a solution  of   (\ref{eq:graz2}) such that  $\bar x(t_0) = x_0 + \Delta x,\, \Delta x = (\xi_1,\xi_2,\ldots,\xi_n),$  and  let $\eta_j$  be the moments of discontinuity of $\bar x(t).$

The following conditions are required in what follows.

\begin{itemize}
\item[(A)] For all $t \in \mathscr{I} \backslash \cup_{i\in\mathscr{A}} \widehat
{(\eta_i,  \theta_i]}, $ the following  equality is satisfied
\begin{eqnarray}
&& \bar x(t)  - x_0(t) = \sum\limits^{n}_{i=1} u_i(t) \xi_i + o(\|\Delta x\|), \label{edd2}
\end{eqnarray}
where $u_i(t) \in PC(\mathscr{I},\theta)$ and $\mathscr{I}$ is a finite subset of $\mathbb R.$

\item[(B)] There exist constants $\nu_{ij}, j\in\mathscr{A},$ such  that
\begin{eqnarray}
&&  \eta_j - \theta_j = \sum\limits^{n}_{i=1} \nu_{ij} \xi_i + o(\|\Delta x\|); \label{edd1}
\end{eqnarray}

\item[$(C)$] The discontinuity moment $\eta_j$ of the near solution  approaches to the discontinuity moment  $\theta_j, j\in\mathscr{A},$ of grazing one as $\xi$ tends to zero.
\end{itemize}

 The   solution $\bar x(t)$   has  a linearization   with  respect   to  solution $x_0(t)$   if   the   condition  $(A)$ is  valid. Moreover,   if   $x_0(\theta_i)$ is   a grazing  point, then the condition $(C)$ is fulfilled  and condition $(B)$ is true if   $x_0(\theta_i)$ is  a   transversal   point. 

 The  solution $x_0(t)$ is $K-$differentiable  with  respect to    the   initial  value $x_0$  on $\mathscr{I},  t_0 \in \mathscr{I},$ if  for  each  solution $\bar x(t)$   with    sufficiently  small  
 $\Delta x$   the   linearization exists. The   functions $u_i(t)$  and   $\nu_{ij}$   depend   on  $\Delta x$   and   uniformly  bounded  on a  neighborhood of  $x_0.$  

\begin{lemma} \cite{akh-kivi2} Assume that the conditions $(H1)$ and $(A2)$ are valid. Then, the function $\tau_i(x)$ is continuous on the set of points near a grazing point which satisfy condition $(N1).$
\end{lemma}

The systems (\ref{half-auto}) and  (\ref{eq:graz2}) are $B-$equivalent, for this reason it is acceptable to linearize system (\ref{eq:graz2}) instead of system (\ref{half-auto}) around $x_0(t)=x(t,t_0,x_0),$ which is a solution of both systems. Thus, by applying linearization to (\ref{eq:graz2}), the system (\ref{lin1}) is obtained. Additionally, the linearization matrix $B_i$ in (\ref{lin1})  for the   grazing point  also has to  be defined by the formula \eqref{grazBi}, where $W_{ix}$ exists by condition $(A2).$

On the basis of the discussion made in Subsections \ref{trans} to \ref{graz1}, one can conclude that the variational system for the solution $x_0(t)$ with the grazing points can be constructed as  a system (\ref{lin1}).

\section{Stability of grazing  periodic  solutions}

Assume additionally  that $f(t,x)$ in \eqref{half-auto}  is $T-$ periodic in time, i.e. $f(t+T,x)=f(t,x),$ for $T>0.$  Let $\Psi(t):\mathbb{R_+}\rightarrow D$ be a periodic solution of (\ref{half-auto})  with   period $T$ and $\theta_i,\ i\in\mathbb{Z},$ be the points of discontinuity satisfying  $\theta_{i+p}=\theta_i + T, p$  is a natural number. 

On the interval $[0,T],$ the periodic solution, $\Psi(t),$ has  $p$ discontinuity moments. Assume that $k$ of them are grazing and the remaining $p-k$ are transversal points.   By means of \eqref{grazBi}, there may appear $2^k$ different values for the matrix $B_i.$ In the following part, to be clear in notations, we will use  $B_i^{(j)},$ where the upper script ${j}$ denotes the number of different matrices and the number ${(j)}$ can take  $2^k$ values at most.

In what follows, we assume  the validity of the next condition.   
 
\begin{itemize}
\item[(A3)] For each $\Delta x \in \mathbb{R}^n,$  the variational system for the near solution $x(t)=x(t,t_0,x_0+\Delta x)$ to $\Psi(t)$ is one of the following $m$ periodic homogeneous linear impulsive systems 

\begin{equation}\label{eq:graz1lin1}
\begin{aligned}
&u^\prime=A(t)u,\\
&\Delta u|_{t=\theta_i}=B_{i}^{(j)} u,
\end{aligned}
\end{equation}
such that   $B_{i+p}^{(j)} =B_{i}^{(j)} ,$ $i\in\mathbb{Z}, \ j=1,\ldots,m,$ where the number $m$ cannot be larger than $ 2^k.$ 
\end{itemize}
We will call the collection of $m$ systems (\ref{eq:graz1lin1})   \textit{the variational system around the periodic grazing orbit}.

So, the variational system (\ref{eq:graz1lin1}) consists of $m$ periodic subsystems.  For each of these systems, we find the matrix of monodromy, $ U_j(T)$ and denote  corresponding  Floquet multipliers  by $\rho_i^{(j)},$ $i=1,\ldots,n,$  $j=1,\ldots,m.$  Next, we need  the following assumption,
\begin{itemize}
\item[(A4)]  $|\rho_i^{(j)}|<1,$ $i=1,\ldots,n,$ for each  $j=1,\ldots,m.$ 
\end{itemize}

\begin{theorem} \cite{akh-kivi2}\label{thmstability} Assume that the conditions $(H1),$ $(A1)-(A4)$ are valid. Then $T-$ periodic solution $\Psi(t)$ of (\ref{half-auto}) is   asymptotically stable.
\end{theorem}
 
\section{ Main result: Regular perturbations}  \label{pertregular}

In the previous part of the paper, we analyze the existence and stability of the periodic solutions of non-
autonomous systems with a stationary impulse condition by applying regular perturbations to the system.
Under certain conditions, the perturbation gives rise the existence of periodic solution in impulsive systems.

To make our investigations, we take into account the following perturbed system:
\begin{equation}\label{half-autoperturbed}
\begin{aligned}
&x^{\prime}=f(t,x)+\mu g(x,\mu),\\
&\Delta x|_{x\in S(\mu)}=I(x )+\mu K(x,\mu),
\end{aligned}
\end{equation}
where $(t,x)\in \mathbb{R}\times G,$ $\mu\in (-\mu_0,\mu_0),$ $\mu_0$ is a fixed positive number. The system \eqref{half-autoperturbed} is $T-$periodic system, i.e.  $f(t+T,x )=f(t,x )$ and $g(t+T,x,\mu )=g(t,x,\mu )$ with some positive number $T.$ Additionally,  $f(t,x)$ is two times differentiable in $x$   and  continuous in time. The function $K(x,\mu)$ is differentiable in $x$ and $\mu $ and $I(x)$ is differentiable in $x.$  The surface of discontinuity of \eqref{half-autoperturbed}, $S(\mu)$ is defined as  $S(\mu)=\{x\in G| \quad \Phi(x )+\mu\phi(x,\mu)=0\},$ where $\Phi(x)  $ second order differentiable in $x$ and $\phi(x,\mu)$ is second and first order differentiable in $x$ and $\mu,$ respectively.

The generating system of \eqref{half-autoperturbed}   is the system \eqref{half-auto}. In  the previous section, we assumed that the generating system has a periodic solution $\Psi(t).$  All assumptions and conditions $(H1),\, (A1)-(A4)$ are also valid in this section. 

Let us seek the periodic solutions of \eqref{half-autoperturbed} around the grazing one. Generally, the investigation on the periodic solutions of such systems  is carried out by utilizing Poincare map which is based on the values of solutions at the period moment. For this reason,   we will analyze the existence of the periodic solution of \eqref{half-autoperturbed} in the light of this map.  But, it may not be differentiable near a grazing point \cite{gbio,Luo1}. To handle with this problem, we make use of the condition $(A1).$  There can be $m$ different linearization systems  around the grazing solution depending on the grazing point. Fix some   $j,$ where $j\in{1,2,\ldots,m}.$  Denote a solution of system \eqref{half-autoperturbed} by
\begin{equation}\label{persol}
x_i^{(j)}(t,\gamma_1^{(j)},\ldots,\gamma_n^{(j)},\mu), \, i=1,2,\ldots,n,
\end{equation}
with initial values 
\begin{equation}\label{persol1}
x_i^{(j)}(0,\gamma_1^{(j)},\ldots,\gamma_n^{(j)},\mu)=\gamma_i^{(j)}, \, i=1,2,\ldots,n.
\end{equation}
Moreover, considering the periodic solution $\Psi(t)=(\Psi_1(t),\ldots,\Psi_n(t))$ of the generating system, it is easy to obtain that
\begin{equation}\label{persol2}
x_i^{(j)}(t,\Psi_1(0),\ldots,\Psi_n(0),0)\equiv\Psi_i(t), \, i=1,2,\ldots,n.
\end{equation}
In order to verify the existence of such periodic solution of \eqref{half-autoperturbed}, it is neccessary and sufficient to check
 the validity of the following equality

\begin{equation}\label{persol3}
\mathscr{P}_i^{(j)}(\gamma_1^{(j)},\ldots,\gamma_n^{(j)})=x_i^{(j)}(T,\gamma_1^{(j)},\ldots,\gamma_n^{(j)},\mu)-\gamma_i^{(j)}, \, i=1,2,\ldots,n.
\end{equation}
By means of the equation \eqref{persol} conditions \eqref{persol1} -\eqref{persol3} are satisfied for $\mu=0,$ $\gamma_i=\Psi_i(0),$ since the generating solution is periodic.

The following conditions for the determinant will be needed for the rest of the paper.
\begin{itemize}
\item[(A5)]\begin{equation}\label{dethalf}\begin{vmatrix} \frac{\partial\mathscr{P}_1^{(j)}(\gamma_1,\ldots,\gamma_n)}{\partial\gamma_1}& \cdots& \frac{\partial\mathscr{P}_1^{(j)}(\gamma_1,\ldots,\gamma_n)}{\partial\gamma_n}\\ \vdots & \ddots& \vdots\\\frac{\partial\mathscr{P}_n^{(j)}(\gamma_1,\ldots,\gamma_n)}{\partial\gamma_1}& \cdots& \frac{\partial\mathscr{P}_n^{(j)}(\gamma_1,\ldots,\gamma_n)}{\partial\gamma_n}  \end{vmatrix}\neq 0.\end{equation}
\end{itemize}

Denote by $I = [a,b]$ an  interval in  $\mathbb{R}.$  Let  $x_1(t)$   be  a piecewise   continuous   function with discontinuity  moments  $\theta^1=\{\theta_i^1\},$ and $x_2$  be  a piecewise   continuous   function with discontinuity  moments  $\theta^2=\{\theta_i^2\}.$

\begin{definition} The function $x_2(t)$ is in the $\epsilon-$neighborhood of $x_1(t)$ on the interval $I$ if 
\begin{itemize}
\item $|\theta_i^1-\theta_i^2|<\epsilon$ for all $\theta_i^1\in I$;
\item the inequality $||x_1(t)-x_2(t)||<\epsilon$ is valid for all t, which satisfy $t\in I\setminus\cup_{\theta_i^1\in I} (\theta_i^1-\epsilon,\theta_i^1+\epsilon).$ 
\end{itemize}
\end{definition}

The topology defined with the help of $\epsilon-$ neighborhoods is called the B-topology. One can easily see that it is Hausdorff and it can be considered also if two functions
$x_1(t)$ and $x_2(t)$ are defined on a semi-axis or on the entire real axis.

 Without loss of generality, assume that the moments of discontinuity of the periodic solution $\Psi(t)$  admits that $0<\theta_1<\ldots<\theta_p<T.$ Let $x^{(j)}(t)=x(t,0,x^{(j)},\mu),  j = 1,\ldots,m,$ be a   periodic solution of the perturbed system (\ref{half-autoperturbed}) with initial values $x^{(j)}(0)=x^{(j)}.$  We   consider $m$  possible  periodic solutions according to  the  number of variational   systems. 
Applying the formulas \eqref{persol}-\eqref{persol3}, the following   equation  can be obtained

\begin{equation}\label{poingraz}
\mathscr{P}^{(j)}(x^{(j)},\mu)= x(t,0,x^{(j)},\mu)- x^{(j)}=0.
\end{equation}
It is satisfied with $x^{(j)}= \Psi(0), \mu =0.$ Now, let us apply implicit function theorem to verify the existence of the periodic solutions of (\ref{half-autoperturbed}) with the help of \eqref{poingraz} in the neighborhood of $(\Psi(0),0).$ For $\mu=0,$ it is easy to obtain the following variational system

\begin{equation}\label{vartheo}
\begin{aligned}
u^\prime=A(t)u,\\
\Delta u(\theta_i)=B_i^{(j)}u(\theta_i),
\end{aligned}
\end{equation}
where $i\in\mathbb{Z}$ and $j=1,\ldots,m.$ The $K-$derivatives of  the solution $x^{(j)}(t)$ in $x^{(j)}$ forms the fundamental matrix $Y^{(j)}(t)$  of the variational system  \eqref{vartheo} with $Y^{(j)}(0)=I,$ where $I$ is the identity matrix. The uniqueness of the periodic solution $\Psi(t)$ implies that 
\begin{equation}\label{poingrazder}
\mathscr{P}_y^{(j)}(\Psi(0),0)=Y^{(j)}(T)-I \neq 0.
\end{equation}
Therefore, the equation \eqref{poingraz} has a unique solution in the neighborhood of $\Psi(0)$ for sufficiently small $|\mu|.$ The suggested periodic solution takes the form $x^{(j)}(t)=x(t,0,x^{(j)}(\mu),\mu),$ where $x^{(j)}(\mu)$ are the initial values of that solution which are obtained uniquely from the equation \eqref{poingraz}.   Now,   we   have to   make the   following   important   observation. Fix a grazing moment $\theta_i,$  and assume that  $B_i^{(j)} = O_n.$   It  is possible that  $x^{(j)}(t)$ intersects  the  surface of discontinuity  near   $\theta_i,$ but  the  linearization  has   been  constructed considering the  opposite  case.   This makes   the  discussion incorrect.  In the  case  $B_i^{(j)} =  W_{ix},$
the suggested   solution, $x^{(j)}(t)$ definitely  intersects  the   surface   of discontinuity   for  small  $|\mu|$  if  one take into   account the   condition $(H1)$  and  continuous   dependence  on the  parameter.   This  is   why  we  consider  the  above discussion only  for  the   single subsystem  of the   variational  equation, when   $B_i^{(j)} =  W_{ix}(\Psi(\theta_i)),$
for  all  $i.$ The   solution  $x^{(j)}(t)$  exists,   and  became closer to $\Psi(0)$ as $\mu$ tends to zero.  Thus, we can conclude that the system (\ref{half-auto}) admits a non-trivial  $T-$ periodic solution, which converge in the $B-$ topology to the $T-$ periodic solution  $\Psi(t)$  of (\ref{half-auto}) as $|\mu|$ tends to zero.

On the basis  of the  above discussion   one can formulate the following assertion.

 \begin{theorem}\label{theopert}Assume that the   conditions $(H1)$ and $(A1)-(A5)$ are valid. Then, there  is  a  non-trivial $T-$periodic solution of (\ref{half-auto}), which converges in the $B-$ topology to the T-periodic solution of (\ref{half-auto}) as $\mu$ tends to zero. 
\end{theorem}

  From the   last   proof  it  implies that  among  possible  $m$ periodic   solutions   only  one  has been  verified. For  all  other  $m-1$ periodic   solutions the  conditions  are not sufficient.  That  is, some additional  research  has to  be done to  prove their existence. If one find theoretical conditions for  that or  a  model with  detailed orbits, then we  can  say   about  bifurcation of periodic   solutions.

\begin{example}
In the previous part of the paper, we analyze the existence and stability of the periodic solutions of non-
autonomous systems with a stationary impulse condition by applying regular perturbations to the system.
Under certain conditions, the perturbation gives rise the existence of periodic solution in impulsive systems. The first mass, $m,$ is subdued impacts with the rigid barrier and the other mass, $M,$ connected to the wall with a spring and damper. The  model  can be observed in Fig.  \ref{mech}.

\begin{figure}[htb]
\centerline{\psfig{figure=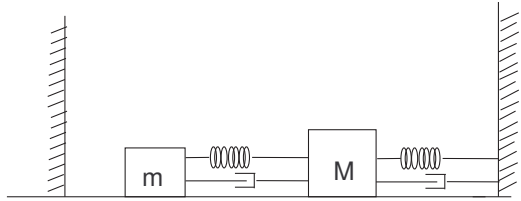,width=7cm}}
\caption{The mechanical model for two degree of freedom oscillator.}
\label{mech}
\end{figure}

The mathematical model for this problem is of the form 
\begin{equation}\label{2dof1}
\begin{aligned}
&mx''+ax^\prime+bx+\epsilon_1(x'-y')+\epsilon_2(x-y)=c\sin(t),\\
&My''+Ay^\prime+\epsilon_1(y'-x')+\epsilon_2(y-x)+B\frac{1}{16}y^3=C,\\
&\Delta x^\prime|_{x\in S}=-(1+R x^\prime) x^\prime ,
\end{aligned}
\end{equation}
where $S=\{(x,x^\prime,y,y^\prime)|\Phi(x,x^\prime,y,y^\prime)=x+1=0\},$ $R$ is the coefficient of restitution which varies between zero and unity.   

The system \eqref{2dof1}  can be normalized as
\begin{equation}\label{2dof1nor}
\begin{aligned}
& x''+\eta_1x^\prime+\eta_2x+\mu_1(x'-y')+\mu_2(x-y)=\eta_3\sin(t),\\
& y''+\xi_1y^\prime+\xi_2y^3-\mu_1(x'-y')-\mu_2(x-y)=\xi_3,\\
&\Delta x^\prime|_{x\in S}=-(1+R x^\prime) x^\prime,
\end{aligned}
\end{equation}
where  $\eta_1=a/m,$ $\eta_2=b/m,$ $\eta_3=c/m,$ $\xi_1=A/M,$ $\xi_2=B/M,$ $\xi_2=C/M,$ $\mu_1=\epsilon_1/m$ and $\mu_2=\epsilon_2/M.$
The generating system corresponding to  normalized system \eqref{2dof1nor} takes the form
\begin{equation}\label{2dof1gen}
\begin{aligned}
& x''+\eta_1x^\prime+\eta_2x=\eta_3\sin(t),\\
& y''+\xi_1y^\prime+\xi_2y^3=\xi_3,\\
&\Delta x^\prime|_{x\in S}=-(1+R x^\prime) x^\prime.
\end{aligned}
\end{equation}

In this model, we will consider the case when the coefficients are $\eta_1=1,$ $\eta_2=1,$ $\eta_3=1,$ $\xi_1=1,$ $\xi_2=1/16$ and $\xi_3=4$ and the coefficient of restitution $R=0.8.$ Defining variables as $x=x_1,$ $x'=x_2,$ $y=x_3$ and $y'=x_4,$ we can obtain that
\begin{equation}\label{2dof1genone}
\begin{aligned}
&x_1'=x_2,\\
&x_2'=-x_2-x_1+\sin(t),\\
&x_3'=x_4,\\
&x_4'=-x_4-\frac{1}{16}x_3^3+4,\\
&\Delta x_2|_{x\in S}=-(1+0.8 x_2) x_2,
\end{aligned}
\end{equation}
where $x=(x_1,x_2,x_3,x_4)$ and $S=\{x|\Phi(x)=x_1+1=0\}.$

It is easy to see that the generating system (\ref{2dof1genone}) consists of two uncoupled oscillators, one of them is an impacting forced spring-mass-damper system and the other one is forced Duffing oscillator. 
The system can be interpreted as the following subequations

\begin{subequations}\label{2dof1genone1}
\begin{equation}\label{2dof1genone1a}
\begin{aligned}
&x_1'=x_2,\\
&x_2'=-x_2-x_1+\sin(t),\\
&\Delta x_2|_{x\in S}=-(1+0.8 x_2) x_2,
\end{aligned}
\end{equation}
and
\begin{equation}\label{2dof1genone1b}
\begin{aligned}
&x_3'=x_4,\\
&x_4'=-x_4-\frac{1}{16}x_3+4,
\end{aligned}
\end{equation}
\end{subequations}

It is easy to verify that the first system, \eqref{2dof1genone1a}, has a $2\pi-$periodic solution $\bar{\Psi}(t)=(-\cos(t),\sin(t))$ which grazes the surface of discontinuity at the moment $t=0$ and at the point $\bar{\Psi}(0)=(-1,0).$ The system \eqref{2dof1genone1b} has a fixed point $(4,0).$  Thus, we have that $\Psi(t)=(-\cos(t),\sin(t),4,0)$ is a solution of the generating system, \eqref{2dof1genone}. In the following part, we will consider   $\Psi(t) $ as   $2\pi-$ periodic  function.   At the moment $t^*=0$   and the point $x^*=(-1,0,4,0) = \Psi(t^*)  $  it  is true  that  $\langle \Phi(x*), f(t*,x*)\rangle=\langle (1,0,0,0), (0,-1,0,0)\rangle=0.$ Now, we can conclude that the point $x^*=(-1,0,4,0)$ is a \textit{grazing point} and  $\Psi(t)$  is a grazing periodic solution.

Because  the two systems, \eqref{2dof1genone1a} and \eqref{2dof1genone1b}, are uncoupled, we will consider the linearization of these systems, separately. 
Let us start with the system  \eqref{2dof1genone1a}.   What   we   can not   do  in this   example,  is to   verify  condition $(A3).$   This  is  a difficult    task, which can be considered for autonomous   planar   systems.   Nevertheless,   it  will  be seen  by  simulations that   a discontinuous   periodic   solution   of the  perturbed   equation exists.   This   is,   because  all   other   sufficient   conditions will  be verified.  Thus,   for  the present  model   we consider two  sorts   of   variational   subsystems   around   $\Psi(t).$   Accordingly, there are two different type of solutions near the periodic solution due to the fact that the periodic solution, $\bar{\Psi}(t),$ is  a grazing one. The first type  is non-impacting and the other is impacting. We will continue with the non-impacting one. For those, the linearization system has the form
  
\begin{equation}\label{2dof1genone1alina}
\begin{aligned}
&u_1'=u_2,\\
&u_2'=-u_2-u_1.
\end{aligned}
\end{equation}
The characteristic multipliers of the system are $\rho_1^{(1)}= 0.0288 + 0.0322i$ and $\rho_2^{(1)}=  0.0288 - 0.0322i.$ All characteristic multipliers are less than unity in magnitude. Now, we can conclude that the periodic solution, $\bar{\Psi}(t)$ is asymptotically stable with respect to inside (non-impacting) solutions.

Next, we will take into account the linearization of \eqref{2dof1genone1a} with respectt to impacting solutions. In the light of the  theoretical   part of the paper the linearization system can be obtained as 
\begin{equation}\label{2dof1genone1alinb}
\begin{aligned}
&u_1'=u_2,\\
&u_2'=-u_2-u_1,\\
&\Delta u(2\pi i)=W_{ix}(\bar{\Psi}(2\pi i))u(2\pi i).
\end{aligned}
\end{equation}
 Considering the formulas \eqref{dert} and \eqref{derw}, the characteristic multipliers of the linearization system, \eqref{2dof1genone1alinb} can be computed as $\rho_1^{(2)}= 0.0868$ and $\rho_2^{(2)}= 0.$ Now, we can say that the characteristic multipliers are inside the unit disc, so the periodic solution is asymptotically stable with respect to impacting (outside) solutions. 

Let us take into account the linearization of system \eqref{2dof1genone1b} at the fixed point, $(4, 0),$ and it is obtained as  
\begin{equation}\label{2dof1genone1alina1}
\begin{aligned}
&u_3'=u_4,\\
&u_4'=-u_4-3u_1.
\end{aligned}
\end{equation}
 The characteristic multipliers of the system are \eqref{2dof1genone1alina1} $\rho_3^{(1)}=0.3679$ and $\rho_4^{(1)}=0.3679.$ Both are inside the unit disc, then it is easy to conclude that the fixed point, $(4, 0),$  is asymptotically stable.

In Fig. \ref{new1}, the grazing periodic solution for system \eqref{2dof1genone1a} is depicted in green. The blue is depicted for the coordinates, $x_1(t)$ and $x_2(t)$ of the system \eqref{2dof1genone} with initial value $(2.12,0, 4, 3)$ and red is simulated for the coordinates, $x_1(t)$ and $x_2(t)$ of the system \eqref{2dof1genone} with initial value $(0,0, 4, 3).$ It is easy to see in Fig. \ref{new1} that both solutions approach the green periodic solution asymptotically, as time increases.

\begin{figure}[htb]
\centerline{\psfig{figure=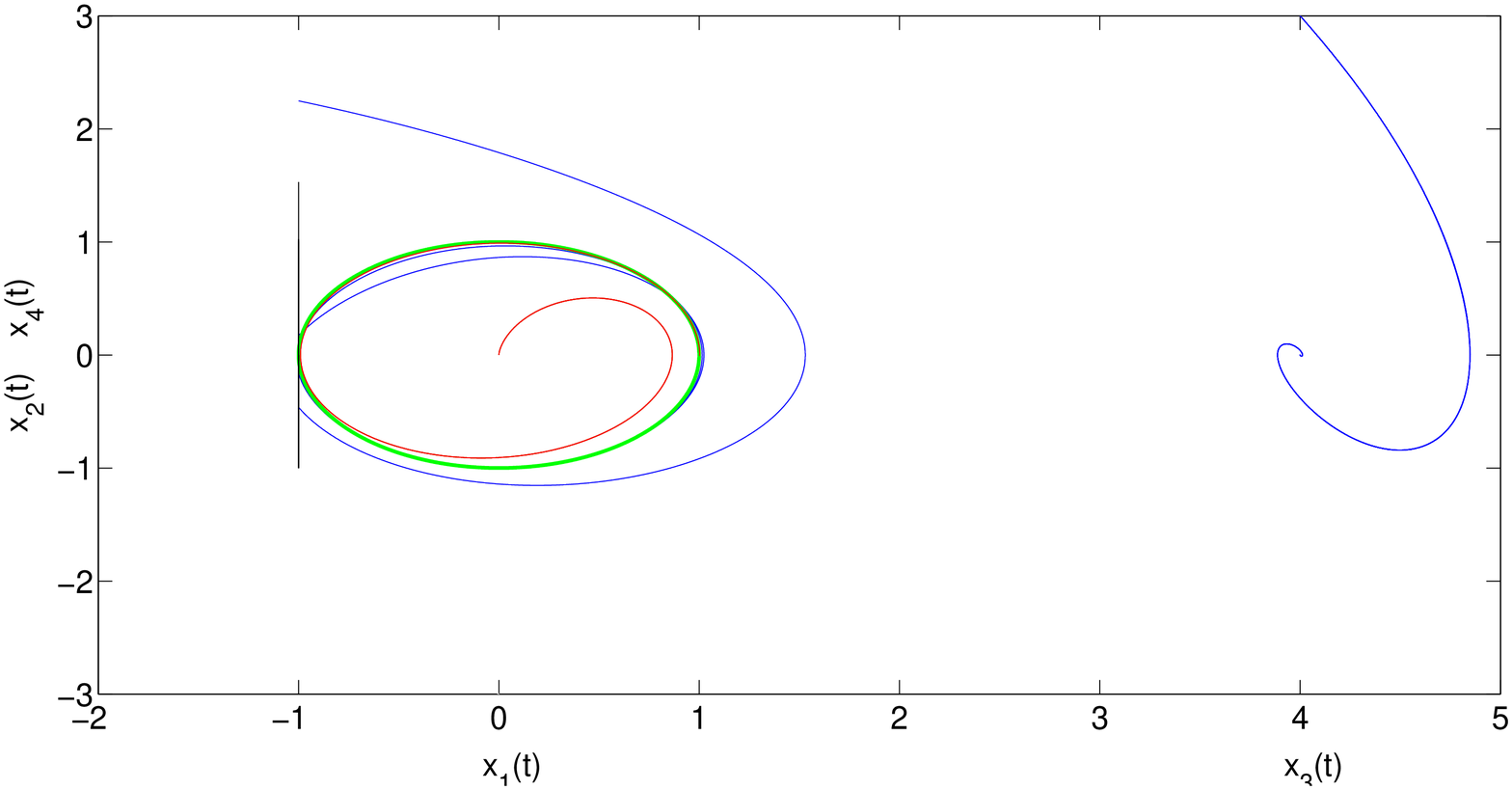,width=9cm}}
\caption{The green is for the grazing periodic solution of  system \eqref{2dof1genone1a}. The blue  and red are for the coordinates, $x_1(t)$ and $x_2(t)$ of the system \eqref{2dof1genone} with initial values $(2.12,0, 4, 3)$ and  $(0,0, 4, 3),$ respectively. }
\label{new1}
\end{figure}

In Figs. \ref{new2} and \ref{new3}, the coordinates $x_1(t),$ $x_2(t)$ and $x_3(t)$ and $x_1(t),$ $x_3(t)$ and $x_4(t)$ of the system \eqref{2dof1genone} are depicted, respectively. Taking into account both of the Figs., we can conclude   geometrically that all near solutions approach to the grazing periodic solution, $\Psi(t),$ of the system \eqref{2dof1genone} as time increases. 

\begin{figure}[htb]
\centerline{\psfig{figure=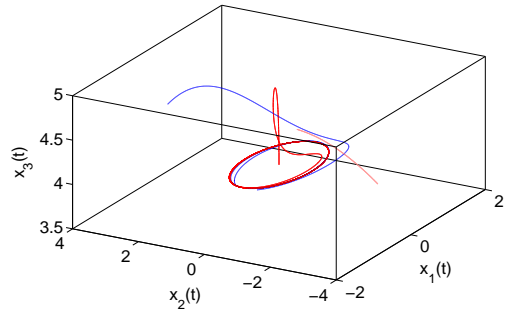,width=7cm}}
\caption{The blue and red are  for the coordinates, $x_1(t),$ $x_2(t)$ and $x_3(t)$ of the system \eqref{2dof1genone} with initial values $(2.12,0, 4, 3)$  and  $(0,0, 4, 3),$ respectively.}
\label{new2}
\end{figure}

\begin{figure}[htb]
\centerline{\psfig{figure=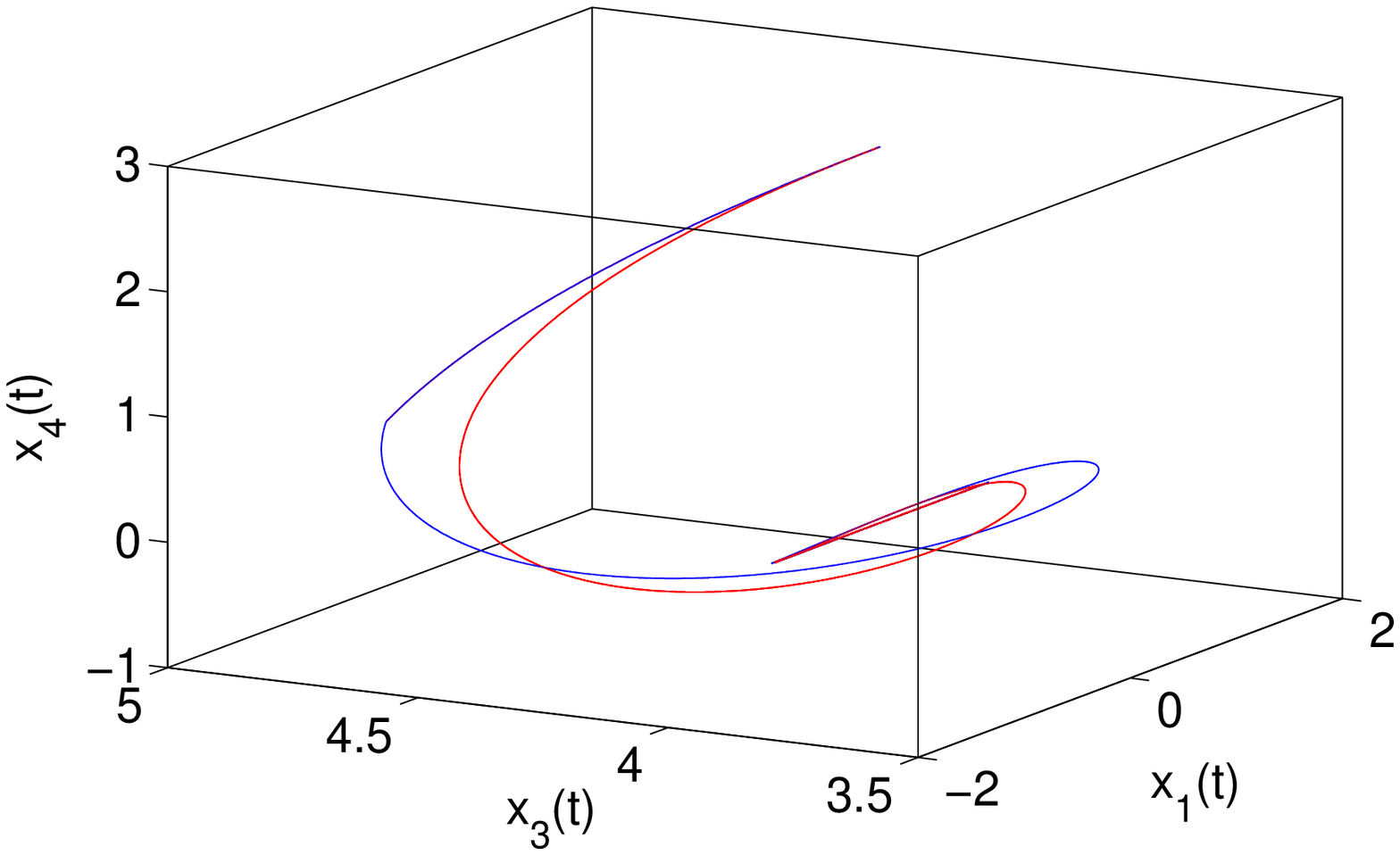,width=7cm}}
\caption{The blue and red are for the coordinates, $x_1(t),$ $x_3(t)$ and $x_4(t)$ of the system \eqref{2dof1genone} with initial values $(2.12,0, 4, 3)$ ans  $(0,0, 4, 3),$  respectively. }
\label{new3}
\end{figure}

Next, our aim to verify that the perturbed system \eqref{2dof1} admits a periodic solution which approaches the periodic solution of the generating system \eqref{2dof1gen} as $\epsilon_1$  and $\epsilon_2$ tend to zero. 
To accomplish it, we will use the formulas and assertions, given in   Section \ref{pertregular}.

Normalizing and defining variables as $x=x_1,$ $x'=x_2,$ $y=x_3$ and $y'=x_4,$ we can obtain that
\begin{equation}\label{2dof1pert}
\begin{aligned}
&x_1'=x_2,\\
&x_2'=-x_2-x_1+\sin(t)-\mu_1(x_4-x_2)-\mu_2(x_1-x_3),\\
&x_3'=x_4,\\
&x_4'=-x_4-\frac{1}{16}x_3^3+4 +\mu_1(x_4-x_2)+\mu_2(x_1-x_3),\\
&\Delta x_2|_{x\in S}=-(1+0.8 x_2) x_2,
\end{aligned}
\end{equation}
where $x=(x_1,x_2,x_3,x_4)$ and $S=\{x|\Phi(x)=x_1+1=0\}.$

 There are two  parameters,  $\mu_1$ and $\mu_2$ which are not single. This does not effect the way of discussion, since the theorem for those systems which have more than one parameter  can be obtained by reformulating  Theorem \ref{theopert}. For this reason, it is worth saying that  applying Theorem \ref{theopert} discussion, it is easy to conclude that the perturbed system \eqref{2dof1} has a periodic solution  and the periodic solution is asymptotically stable. It can be observed in Figs \ref{figpert1}, \ref{figpert2}.

\begin{figure}[htb]
\centerline{\psfig{figure=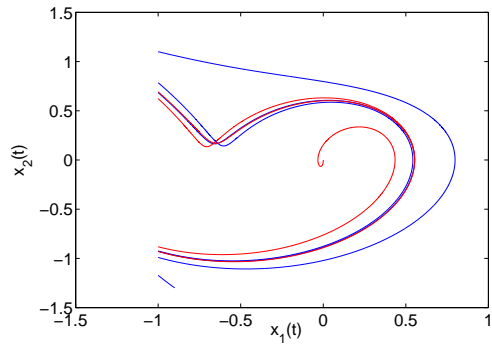,width=7cm}}
\caption{The blue and red are for the coordinates, $x_1(t),$ $x_2(t)$  for the solution of system \eqref{2dof1pert} with initial values $(1,0, 4.15, 0.1)$ and  $(0,0, 4.15, 0),$  respectively. }
\label{figpert1}
\end{figure}
 
\begin{figure}[htb]
\centerline{\psfig{figure=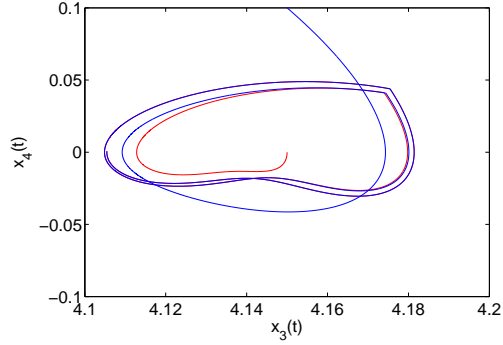,width=7cm}}
\caption{The blue and red are for the coordinates, $x_3(t),$ $x_4(t)$  for the solution of system \eqref{2dof1pert} with initial values $(1,0, 4.15, 0.1)$ and  $(0,0, 4.15, 0),$  respectively.}
\label{figpert2}
\end{figure}

\end{example}

\begin{example}
Next, we will consider the following bi-laterally connected oscillator with    a  non-trivial four-dimensional periodic solution. Let us consider the following system  
\begin{equation}\label{2dof1duff}
\begin{aligned}
&x''+x^\prime+x+\mu_1(x'-y')+\mu_2(x-y)=\sin(t),\\
&y''+2y^\prime+y^3+\mu_1(y'-x')+\mu_2(y-x)+\frac{1}{16}y^3+\sin(t)=4,\\
&\Delta x^\prime|_{x\in S}=-(1+0.8 x^\prime) x^\prime ,
\end{aligned}
\end{equation}
where $x=(x,x^\prime,y,y^\prime)$ $S=\{x|\Phi(x)=x=-1\}.$ The generating system for  \ref{2dof1duff} is of the form 
\begin{equation}\label{2dof1duffgen}
\begin{aligned}
&x''+x^\prime+x=\sin(t),\\
&y''+y^\prime+\frac{1}{16}y^3=4+\sin(t),\\
&\Delta x^\prime|_{x\in S}=-(1+0.8 x^\prime) x^\prime ,
\end{aligned}
\end{equation}
Defining the variables $x=x_1,$ $x'=x_2,$ $y=x_3$ and $y'=x_4,$ the systems \eqref{2dof1duffgen} and  \eqref{2dof1duff}  can be rewritten as
\begin{equation}\label{2dof1duffgenone}
\begin{aligned}
&x_1'=x_2,\\
&x_2'=-x^\prime-x-\sin(t),\\
&x_3'=x_4,\\
&x_4'=-x_4-\frac{1}{16}x_3^3+4+\sin(t),\\
&\Delta x_2|_{x\in S}=-(1+0.8 x_2) x_2,
\end{aligned}
\end{equation}
and 
\begin{equation}\label{2dof1duffpertone}
\begin{aligned}
&x_1'=x_2,\\
&x_2'=-x^\prime-x-\sin(t)-\mu_1(x_4-x_2)-\mu_2(x_3-x_1),\\
&x_3'=x_4,\\
&x_4'=-x_4-\frac{1}{16}x_3^3+\mu_1(x_4-x_2)+\mu_2(x_3-x_1)+4+\sin(t),\\
&\Delta x_2|_{x\in S}=-(1+0.8 x_2) x_2,
\end{aligned}
\end{equation}

where $x=(x_1,x_2,x_3,x_4)$ and $S=\{x|\Phi(x)=x_1=-1\}.$ 

The   non-perturbed  system \eqref{2dof1duffgenone} consists of two uncoupled oscillators and they are:
\begin{subequations}
\begin{equation}\label{2dof1duffgenonea}
\begin{aligned}
&x_1'=x_2,\\
&x_2'=-x^\prime-x-\sin(t),\\
&\Delta x_2|_{x\in S}=-(1+0.8 x_2) x_2,
\end{aligned}
\end{equation}
\begin{equation}\label{2dof1duffgenoneb}
\begin{aligned}
&x_3'=x_4,\\
&x_4'=-x_4-\frac{1}{16}x_3^3+4+\sin(t).
\end{aligned}
\end{equation}
\end{subequations}

The system \eqref{2dof1duffgenonea} has a grazing cycle which is of the form $\bar{\Psi}(t)=(-\cos(t),\sin(t)).$    Since of the system \eqref{2dof1genone1b} has an equilibrium with characteristic exponents having negative real part, by applying theorems for the existence of solutions of quasi-linear systems \cite{ref1f,minorski}, one can prove that the periodic solution of the system exists, whenever the coefficient of $x_3^3$ in the system is sufficiently small. Taking into account the coefficient $1/16$ we obtain   that   there is  a $2\pi$-periodic  asymptotically  stable  solution for that system, which can be seen in the Fig.  \ref{figduff2}. Let us denote the $2\pi-$periodic solution of  \eqref{2dof1duffgenone} by $\Psi(t)(-\cos(t),\sin(t),\Psi_3(t),\Psi_4(t)).$ Utilizing the formula $\langle \Phi(x^*),f(t^*,x^*)\rangle,$ at the point $(t^*,x^*)=(0,\Psi(0))=(0,-1,0,\Psi_3(0),\Psi_4(0)),$ it is easy to verify that the point $x^*$ and the moment $t^*$ are grazing. So, the periodic solution, $\Psi(t),$ is a grazing one.

Figs. \ref{figduff3} and \ref{figduff4} are depicted to show the asymptotic properties of the periodic    solution in $3-$dimensional space.

\begin{figure}[htb]
\centerline{\psfig{figure=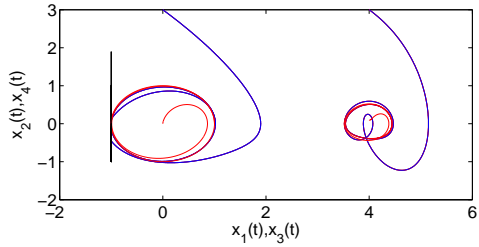,width=7cm}}
\caption{On the left part, the blue and red are for the coordinates, $x_1(t),$ $x_2(t)$  for the solution of system \eqref{2dof1duffgenone} with initial values $(1,0, 4.15, 0.1)$ and  $(0,0, 4.15, 0),$  respectively.
On the right part, the blue and red are for the coordinates, $x_3(t),$ $x_4(t)$  for the solution of system \eqref{2dof1duffgenone} with initial values $(1,0, 4.15, 0.1)$ and  $(0,0, 4.15, 0),$  respectively.}
\label{figduff2}
\end{figure}

\begin{figure}[htb]
\centerline{\psfig{figure=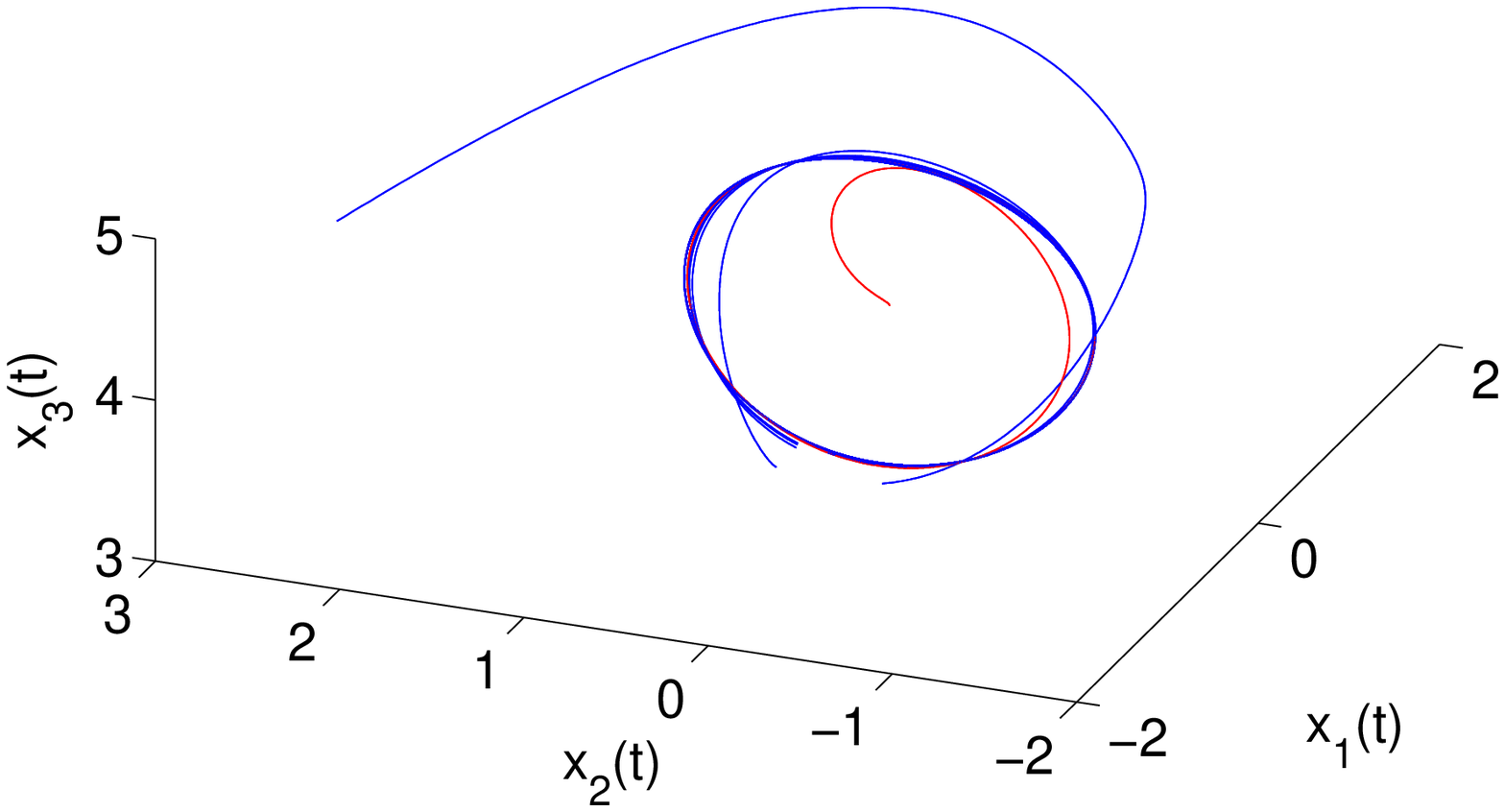,width=7cm}}
\caption{The blue and red are for the coordinates, $x_1(t),$ $x_2(t)$ $x_3(t),$ for the solution of system \eqref{2dof1duffgenone} with initial values $(1,0, 4.15, 0.1)$ and  $(0,0, 4.15, 0),$  respectively. }
\label{figduff3}
\end{figure}
 
\begin{figure}[htb]
\centerline{\psfig{figure=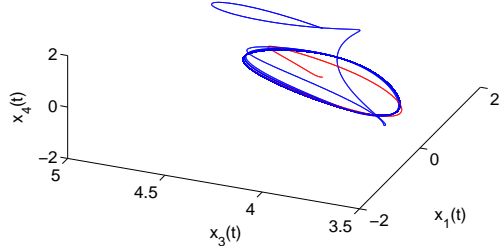,width=7cm}}
\caption{The blue and red are for the coordinates,$x_1(t),$ $x_3(t),$ $x_4(t)$  for the solution of system \eqref{2dof1duffgenone} with initial values $(1,0, 4.14, 0.1)$ and  $(0,0, 4.15, 0),$  respectively.}
\label{figduff4}
\end{figure}

Next,  we   consider   simulations   for the  perturbed   discontinuous periodic   solution. In both Figs. \ref{figpertduff2}, the outside and inside solutions which are drawn in blue and red, respectively approach the periodic solution of the system  \eqref{2dof1duffpertone}, as time increases. The Figs. \ref{figpertduff3} and \ref{figpertduff4} in drawn for the coordinates $x_1(t),$ $x_2(t)$ $x_3(t),$ and $x_1(t),$ $x_3(t),$ $x_4(t),$ respectively. From that figures, one can see the asymptotic properties of the solutions of the perturbed system \eqref{2dof1duffpertone}. In order to obtain better view, if one project the Figs. \ref{figpertduff3} and \ref{figpertduff4} into $x_1-x_2$  and $x_3-x_4$ planes, respectively, one can obtain the left and right parts of Fig. \ref{figpertduff2}, respectively.

\begin{figure}[htb]
\centerline{\psfig{figure=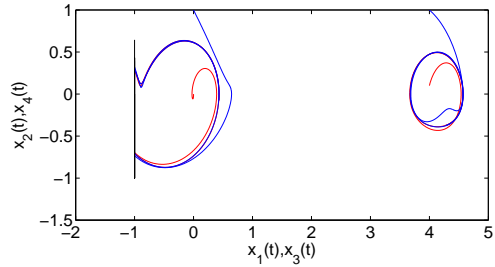,width=7cm}}
\caption{In the left, the blue and red are for the coordinates, $x_1(t),$ $x_2(t)$  for the solution of system \eqref{2dof1duffpertone} with initial values $(1,0, 4.15, 0.1)$ and  $(0,0, 4.15, 0),$  respectively. In the right, the blue and red are for the coordinates, $x_3(t),$ $x_4(t)$  for the solution of system \eqref{2dof1duffpertone} with initial values $(1,0, 4.15, 0.1)$ and  $(0,0, 4.15, 0),$  respectively.}
\label{figpertduff2}
\end{figure}

\begin{figure}[htb]
\centerline{\psfig{figure=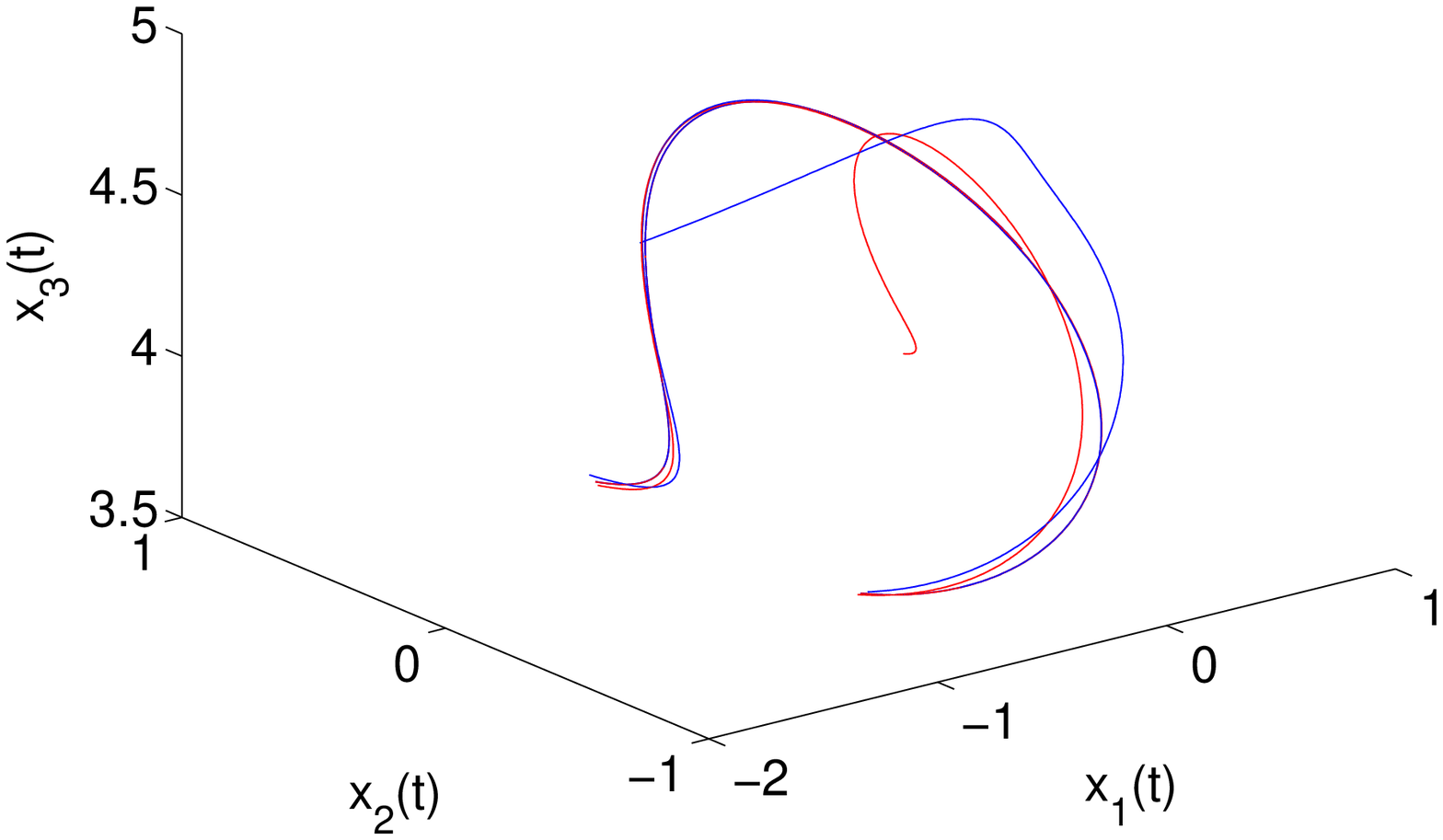,width=7cm}}
\caption{The blue and red are for the coordinates, $x_1(t),$ $x_2(t)$ $x_3(t),$ for the solution of system \eqref{2dof1pert} with initial values $(1,0, 4.15, 0.1)$ and  $(0,0, 4.15, 0),$  respectively. }
\label{figpertduff3}
\end{figure}
 
\begin{figure}[htb]
\centerline{\psfig{figure=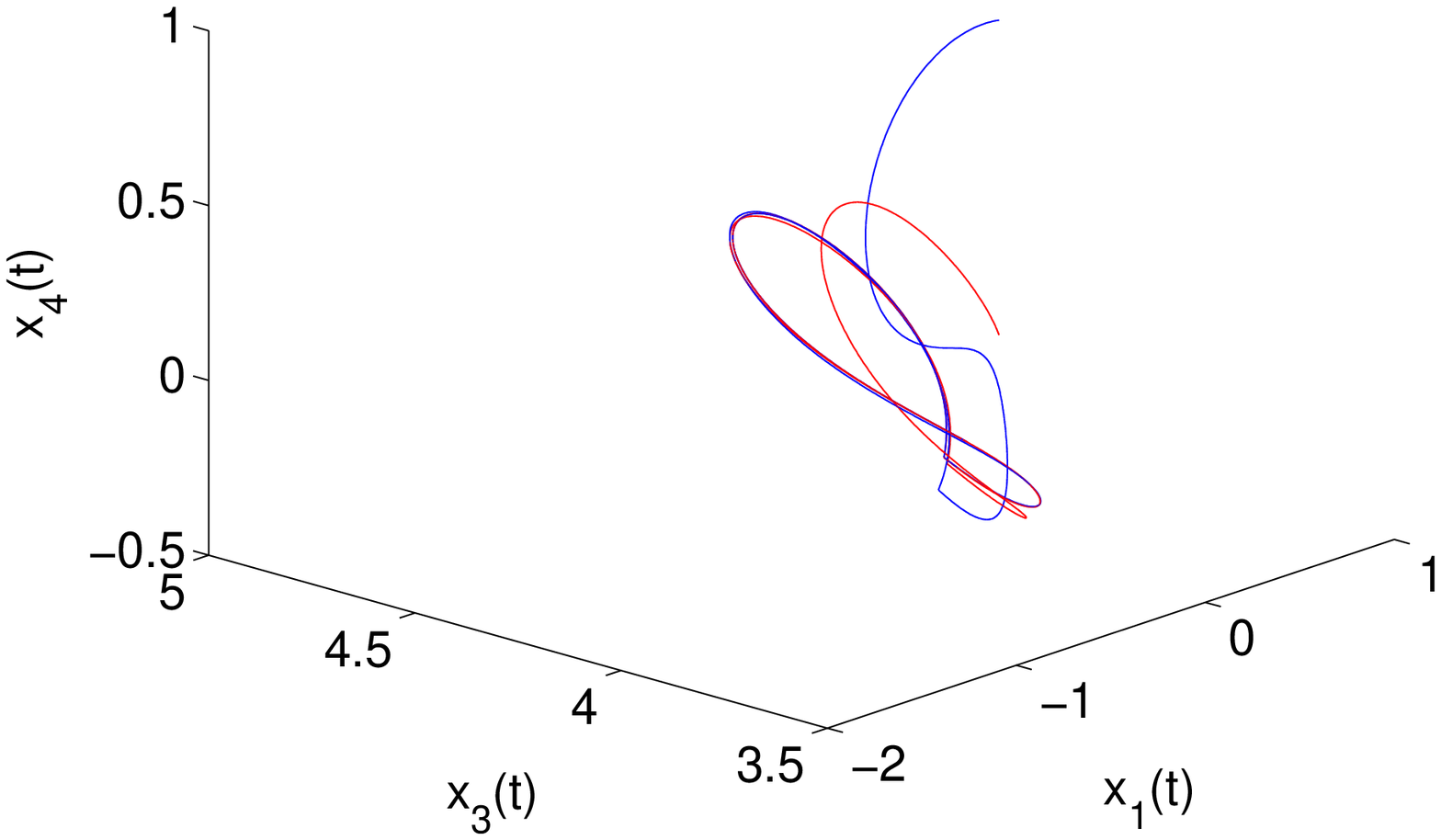,width=7cm}}
\caption{The blue and red are for the coordinates,$x_1(t),$ $x_3(t),$ $x_4(t)$  for the solution of system \eqref{2dof1pert} with initial values $(1,0, 4.14, 0.1)$ and  $(0,0, 4.15, 0),$  respectively.}
\label{figpertduff4}
\end{figure}

\end{example}

\section{Discussion}

The singularity provided by grazing which appears  in the Poincare map, if one want to analyze the problem of stability the following mapping approaches  such as zero time discontinuity mapping \cite{budd1} and Nordmark mapping \cite{Nord1}-\cite{Nord2006} should be considered.  In the literature, as distinct from the mapping results, Ivanov \cite{I1},\cite{I2} analyzed the stability of the grazing periodic solution  half autonomous systems under a parameter variation in the vector field through the variational system approach. As different from the theoretical results, the singularity in our analysis appears during the construction of linearization at the moment of discontinuity. By harmonizing the vector field, the  barrier and the jump equations, the singularity is suppressed in the system.

We provide some examples with simulations to demonstrate the practicability of our theoretical results. In addition, this work can be applied the integrate and fire neuron models which intersects the threshold tangentially. We propose that such phenomenon can be understood as the activity of a neuron cell transfers to the non-firing stage to firing stage.

By applying regular perturbations to the half autonomous system, we investigate the existence of periodic solution of the perturbed system. We derive rigorous mathematical method for the analysis of discontinuous trajectories near grazing orbits. If there is no impacts  in  models,  our results can be easily reduced to those for finite dimensional continuous dynamics. That is why, this method is convenient to investigate infinite dimensional problems and periodic solutions of functional differential equations and bifurcation theory.

In general, the analysis of periodic solutions by using implicit function theorem is not applicable in systems which have graziness. Because grazing point may violate the differentiability of the Poincare map. For this in literature many methods have been used   such as Nordmark map \cite{Nord1}-\cite{Nord2006} and zero time discontinuity mapping (ZDM) \cite{budd1}. By using special assumptions, we investigate the existence and stability of periodic solution of the perturbed system without disrupting the nature of the mechanisms with impacts.


\section{References}

\begin{thebibliography}{30}

\bibitem{prii1}   Piiroinen PT,   Virgin LN,  Champneys AR. Chaos and Period-Adding: Experimental and Numerical Verification of the Grazing Bifurcation.  J. Nonlinear Sci. 2004;14:383–404.

\bibitem{prii2} Eyres RD, Piiroinen PT, Champneys AR, and  Lieven NAJ, Grazing Bifurcations and Chaos in the Dynamics of a Hydraulic Damper with Relief Valves. SIAM J Appl Dyn Syst 2005; 4: 1076-1106.

\bibitem{nordprii1} Nordmark, AB and Piiroinen PT, Simulation and stability analysis of impacting systems with complete chattering. Nonlinear Dynam 2009; 58(1):85-106.

\bibitem{kry11}  Kryzhevich SG, Chaos in vibroimpact systems with one degree of freedom in a neighborhood of chatter generation. Diff Equat 2010;   46(10):1409–1414.

\bibitem{Bernardo-Hogan2010} di Bernardo M, Hogan SJ. Discontinuity-induced bifurcations of piecewise smooth dynamical systems. Philos T Roy Soc A 2010;368:4915-4935. 


\bibitem{budd2001}  di Bernardo M,  Budd CJ,   Champneys AR. Grazing bifurcations in n-dimensional piecewise-smooth dynamical systems.  Physica D 2001;160:222–254.


\bibitem{budd1998} di  Bernardo M,  Budd CJ, Champneys AR. Grazing, skipping and sliding: analysis of the nonsmooth dynamics of the DC/DC buck converter. Nonlinearity 1998;11:858–890.


\bibitem{budd1} Budd CJ. Non-smooth dynamical systems and the grazing bifurcation. Nonlinear Mathematics and its Applications, Guildford, Cambridge Univ. Press; 1996: 219-235.
 
\bibitem{Chin} Chin W, Ott E, Nusse HE, Grebogi C. Universal behavior of impact oscillators near grazing incidence. Phys Letters A 1995;201(2):197-204.


\bibitem{gbio} Chin W, Ott E, Nusse HE, Grebogi C. Grazing bifurcations in impact oscillators.  Phys Rev E 1994; 50:4427-4444.



\bibitem{Hosa-Champneys}   H$\ddot{o}$s C,    Champneys AR. Grazing bifurcations and chatter in a pressure relief valve model.  Physica D 2012;241:2068–2076.

\bibitem{fech} Feckan M,  Bifurcation and Chaos in Discontinuous and Continuous Systems.   Springer$\&$HEP, Berlin; 2011.


\bibitem{feigin70}   Feigin MI. Doubling of the oscillation period with C-bifurcations in piecewise continuous systems. J Appl Math Mech-USS 1970;34:861–869.


\bibitem{I1} Ivanov AP. Impact oscillations: linear theory of stability and bifurcations. J Sound Vib 1994;178: 361-378.

\bibitem{I2} Ivanov  AP. Bifurcations in impact systems.  Chaos Soliton Fract 1996;7:1615-1634.



\bibitem{analysis} Ing  J, Pavlovskaia E, Wiercigroch M, Banerjee S. Bifurcation analysis of an impact oscillator with a one-sided elastic constraint near grazing. Physica D  2010;239(6):312-321.

 
\bibitem{kry}  Kryzhevich SG. Grazing bifurcation and chaotic oscillations of vibro-impact systems with one
degree of freedom.  J Appl Math Mech-USS 2008; 72:383-390.


\bibitem{www} Kryzhevich S,  Wiercigroch M.  Topology of vibro-impact systems in the neighborhood of grazing.  Physica D: Nonlinear Phenomena  2012;241(22):919-1931.


\bibitem{fraggraz} Luo ACJ.  On grazing and strange attractors fragmentation in non-smooth dynamical systems. Commun Nonlinear Sci Numer Simulat2006;11(8):922-933.


\bibitem{Luo2006}  Luo ACJ. Singularity and Dynamics on Discontinuous Vectorv fields. Elsevier; 2006.

\bibitem{luotime}  Luo ACJ.  Discontinuous Dynamical Systems on Time-varying Domains. HEP; 2009.



\bibitem{Luo2005a}   Luo ACJ. A theory for non-smooth dynamical systems on connectable domains. Commun Nonlinear Sci Numer Simulat 2015;10:1-55.

\bibitem{multi} Mats HF. Grazing bifurcations in multibody systems, Nonlinear Anal-Theor 1997;30(7): 4475-4483.

\bibitem{Nord1}  Nordmark AB. Non-periodic motion caused by grazing incidence in an impact oscillator. J Sound Vib 1991;145:279-297.

\bibitem{Nord97} Nordmark AB. Universal limit mapping in grazing bifurcations. Phys Rev E 1997;55:266-270.

\bibitem{Nord2001}Nordmark AB.   Existence of periodic orbits in grazing bifurcations of impacting mechanical oscillators.  Nonlinearity 2001;14:1517–1542.
 
\bibitem{Nord2006}   Nordmark AB, Kowalczyk PA. Codimension-two scenario of sliding solutions in grazing-sliding bifurcations. Nonlinearity 2006;19(1):1–26.


\bibitem{shaw-holmes} Shaw SW, Holmes PJ. A periodically forced piecewise linear oscillator. J Sound Vib   1983;90(1):129-155.


\bibitem{winston} Whiston  GS. Singularities in vibro-impact dynamics. J Sound Vib 1992;152:427-460.


\bibitem{ref1}  Akhmet M.  Principles of Discontinuous Dynamical Systems. Springer-Verlag, Berlin, Heidelberg; 2010.


\bibitem{akh-kivi}  Akhmet M, K{\i}v{\i}lc{\i}m A. Discontinuous dynamics with grazing points. Commun Nonlinear Sci Numer Simulat  2016;38:218-242.

\bibitem{akh-kivi2} Akhmet M, K{\i}v{\i}lc{\i}m A. Stability in non-autonomous periodic systems with grazing
stationary impacts.  Carpathian J Math, (accepted).

\bibitem{brogliato}  Brogliato B. Impacts in Mechanical Systems.  Springer-Verlag, Berlin, Heidelberg; 2000. 


\bibitem{impactdef} Akhmet M, K{\i}v{\i}lc{\i}m A.   The Models with Impact Deformations.  Discontinuity, Nonlinearity, and Complexity 2015;4:49-78. 


\bibitem{Hartmanode} Hartman P. Ordinary Differential Equations, SIAM; 2002.


\bibitem{Luo1} Luo G, Xie J,  Zhu X, Zhang J.  Periodic motions and bifurcations of a vibro-impact system. Chaos Soliton Fract 2008;36:1340–1347.

\bibitem{ref1f} Farkas M. Periodic Motions. Springer-Verlag; 1994. 

\bibitem{minorski} 	 Minorsky N.  Nonlinear oscillations. 	Malabar, Fla. Krieger; 1987.



\end{thebibliography}
\end{document}